\documentclass[conference]{IEEEtran}
\IEEEoverridecommandlockouts
\usepackage{cite}
\usepackage{amsmath,amssymb,amsfonts}
\usepackage{algorithmic}
\usepackage{graphicx}
\ifCLASSOPTIONcompsoc
    \usepackage[caption=false, font=normalsize, labelfont=sf, textfont=sf]{subfig}
\else
\usepackage[caption=false, font=footnotesize]{subfig}
\fi
\usepackage{textcomp}
\usepackage{xcolor}
\usepackage{lipsum}

\def\BibTeX{{\rm B\kern-.05em{\sc i\kern-.025em b}\kern-.08em
    T\kern-.1667em\lower.7ex\hbox{E}\kern-.125emX}}
\catcode`\@=11
\newcommand{\gsim}{${\mathrel{\mathpalette\@versim>}}$}
\newcommand{\lsim}{${\mathrel{\mathpalette\@versim<}}$}
\newcommand{\@versim}[2]{\lower 2.9truept \vbox{\baselineskip 0pt \lineskip
    0.5truept \ialign{$\m@th#1\hfil##\hfil$\crcr#2\crcr\sim\crcr}}}
\catcode`\@=12

\newcommand{\ddeg}{$^{o}$}

\def\be{\begin{equation}}
\def\bea{\begin{eqnarray}}
\def\ee{\end{equation}}
\def\eea{\end{eqnarray}}

\begin{document}

\title{The Next Generation Arecibo Telescope: A preliminary study
\thanks{Invited paper for the ICEAA-IEEE APWC conference, Venice, Italy, Oct 9-13, 2023}
}

\author{\IEEEauthorblockN{D. Anish Roshi}
\IEEEauthorblockA{\textit{Arecibo Observatory} \\
\textit{University of Central Florida}\\
Orlando, USA \\
0000-0002-1732-5990}
\and
\IEEEauthorblockN{Sean Marshall}
\IEEEauthorblockA{\textit{Arecibo Observatory} \\
\textit{University of Central Florida}\\
Arecibo, USA \\
0000-0002-1732-5990}
\and
\IEEEauthorblockN{Amit Vishwas}
\IEEEauthorblockA{\textit{Department of Astronomy} \\
\textit{Cornell University}\\
Ithaca, USA \\
0000-0002-4444-8929}
\and
\IEEEauthorblockN{Mike Sulzer}
\IEEEauthorblockA{\textit{Arecibo Observatory} \\
\textit{University of Central Florida}\\
Arecibo, USA \\
Michael.Sulzer@ucf.edu}
\and
\IEEEauthorblockN{P. K. Manoharan}
\IEEEauthorblockA{\textit{Arecibo Observatory} \\
\textit{University of Central Florida}\\
Arecibo, USA \\
0000-0003-4274-211X}
\and
\IEEEauthorblockN{Maxime Devogele}
\IEEEauthorblockA{\textit{Arecibo Observatory} \\
\textit{University of Central Florida}\\
Arecibo, USA \\
mdevogele@ucf.edu}
\and
\IEEEauthorblockN{Flaviane Venditti}
\IEEEauthorblockA{\textit{Arecibo Observatory} \\
\textit{University of Central Florida}\\
Arecibo, USA \\
0000-0001-9150-8376}
\and
\IEEEauthorblockN{Allison Smith}
\IEEEauthorblockA{\textit{Arecibo Observatory} \\
\textit{University of Central Florida}\\
Arecibo, USA \\
0000-0002-4772-9670}
\and
\IEEEauthorblockN{Sravani Vaddi}
\IEEEauthorblockA{\textit{Arecibo Observatory} \\
\textit{University of Central Florida}\\
Arecibo, USA \\
0000-0003-3295-6595}
\and
\IEEEauthorblockN{Arun Venkataraman}
\IEEEauthorblockA{\textit{Arecibo Observatory} \\
\textit{University of Central Florida}\\
Arecibo, USA \\
arun@naic.edu}
\and
\IEEEauthorblockN{Phil Perillat}
\IEEEauthorblockA{\textit{Arecibo Observatory} \\
\textit{University of Central Florida}\\
Arecibo, USA \\
phil@naic.edu}
\and
\IEEEauthorblockN{Julie Brisset}
\IEEEauthorblockA{\textit{Florida Space Institute} \\
\textit{University of Central Florida}\\
Orlando, USA \\
Julie.Brisset@ucf.edu}
}

\maketitle

\begin{abstract}
The Next Generation Arecibo Telescope (NGAT) was a concept presented in a white paper \cite{b1} developed by members of the Arecibo staff and user community immediately after the collapse of the 305~m legacy telescope. A phased array of small parabolic antennas placed on a tiltable plate-like structure forms the basis of the NGAT concept. The phased array would function both as a transmitter and as a receiver. This envisioned state of the art instrument would offer capabilities for three research fields, viz. radio astronomy, planetary and space \& atmospheric sciences. The proposed structure could be a single plate or a set of closely spaced segments, and in either case it would have an equivalent collecting area of a parabolic dish of size 300~m. In this study we investigate the feasibility of realizing the structure. Our analysis shows that, although a single structure $\sim$300~m in size is achievable, a scientifically competitive instrument 130 to 175~m in size can be developed in a more cost effective manner. We then present an antenna configuration consisting of one hundred and two 13~m diameter dishes. The diameter of an  equivalent collecting area single dish would be $\sim$130~m, and the size of the structure would be $\sim$146~m. The weight of the structure is estimated to be 4300~tons which would be 53\% of the weight of the Green Bank Telescope. We refer to this configuration as NGAT-130. We present the performance of the NGAT-130 and show that it surpasses all other radar and single dish facilities. Finally, we briefly discuss its competitiveness for radio astronomy, planetary and space \& atmospheric science applications.  
\end{abstract}


\section{Introduction}
\label{sec:int}

The Arecibo Observatory (AO) hosted the most sensitive radio telescope and the most powerful radar system until the unexpected collapse of the 305~m “legacy” telescope on December 1, 2020. The facility served three scientific communities: the planetary science, the space and atmospheric sciences, and the radio astronomy communities. Its collapse has produced a significant void in these scientific fields. Any future multidisciplinary facility should enable cutting-edge capabilities for all three scientific fields. 

Immediately after the collapse of the legacy telescope, members of AO staff and the three scientific communities had extensive discussions regarding the rebuild of a future facility. The outcome of these discussions can be summarized as follows \cite{b1}. Any new facility must meet the capability requirements listed below: 
\begin{itemize}
    \item Planetary science: 5~MW of continuous wave transmitting power at 2 to 6~GHz, 1 to 2~arcmin beam width at these frequencies, and increased sky coverage.
    \item Atmospheric science: 0\ddeg\ to 45\ddeg\ sky coverage from zenith to observe both parallel and perpendicular directions to the geomagnetic field, 10~MW peak transmitting power at 430~MHz (also at 220~MHz under consideration) and excellent surface brightness sensitivity.
    \item Radio astronomical science: Excellent sensitivity over 200~MHz to 30~GHz frequency range, increased sky coverage and telescope pointing up to 48\ddeg\ from zenith to observe the Galactic Center.
    \item The initial suggestion was to have a total collecting area for the facility equivalent to a parabolic dish of diameter 300~m.
\end{itemize}

In order to accomplish the above scientific goals, we presented an innovative concept of a compact, phased array of dishes on a steerable plate-like structure. This concept is referred to as the Next Generation Arecibo Telescope (NGAT). This new concept requires extensive engineering studies. Here we present a preliminary study to obtain an estimate for the size and weight of the structure. The study is based on the `von Hoerner Model' \cite{b2},  a cost effective solution to build a large, steerable `single reflector' telescopes (see for example \cite{b7}). In Section~\ref{sec:model}, we discuss the model, the constraints we imposed for the study, and present the results. Based on this study, we present in Section~\ref{sec:ngat130} a new, cost effective, and scientifically competitive configuration, referred to as NGAT-130. A concise list of unique scientific usage of the NGAT-130 is given in Section~\ref{sec:sci} and future work is discussed in Section~\ref{sec:future}.
 
\section{von Hoerner Model}
\label{sec:model}

The von Hoerner model \cite{b2} simplifies a steerable telescope structure to an octahedron. 
The octahedron would be supported at two points and turned from a third point to make the structure steerable. The reflector surface would be constructed on the flat, `square' shaped surface in its center with the focus  located perpendicular to this surface on the fourth point of the octahedron. As the telescope tilts, the structure would deform under its own gravity. Moreover, deflections due to wind and temperature gradient would be superposed on this structural deformation. The octahedral structure provides minimum deflection from its own weight and from wind. Given such a structure and the assumed construction of the reflector surface on the structure, \cite{b2} provides expressions to estimate the weight of the structure as well as relationships between the root mean square (RMS) deflections on the structure due to gravity, wind and temperature gradient and its diameter. 

The NGAT concept consists of a planar array of small parabolic dishes placed on a supporting structure that could be tilted to point the array to different directions in the sky. Each dish has its own focal point and so the ``fourth'' point of the octahedron would not be required. Removing the four legs connecting the ``fourth'' point reduces the weight of the structure but the magnitude of the deflection would be different from those of a symmetric octahedron. Nevertheless, we apply the von Hoerner model here to get an estimate for the weight and size of a realizable NGAT structure. To apply the von Hoerner model, we imagine replacing the framework supporting the reflector surface with an equivalent framework to support the array of dishes, their individual parabolic surfaces, and the electronic and cooling components required for each dish. Unlike in conventional parabolic telescopes, the framework for the NGAT would be constructed so as to support a planar array of dishes. 

We are interested in exploring a NGAT structure with diameter greater than $\sim$100~m. For such a telescope size, the wind and thermal deflections are small compared to that due to gravity if the structure is build at least for survival condition \cite{b2}. The survival wind speed considered is $\sim$136~mph -- 5.4 times more than the typical operating wind speed ($\sim$25~mph) of the telescope. We make the following assumptions to obtain the weight and size estimates for the NGAT structure.
\begin{itemize}
\item 
We assume that the magnitude of gravitational deflection is roughly similar to that for the structure modeled by \cite{b2}. 

\item 
The changes in gravitational deflection are slow compared to the telescope tracking time and the spatial scale of the deflection is larger than the size of the small dishes. Thus the gravitational deflection can be compensated by re-pointing and focusing the optics of the dishes. This assumption is justified by routine application of the surface adjustments made on the Green Bank Telescope (GBT) at frequencies $>$ 20~GHz to recover the loss in aperture efficiency due to gravitational deformation\cite{b3}. 
\end{itemize}

 The weight of the telescope structure can be divided into two parts: (1) active weight and (2) passive weight. The active weight would be that of the load-bearing part of the structure which opposes deflections. For conventional telescopes, the passive weight includes those of the parabolic surface, support framework and parts of the drive mechanism. The passive weight of the NGAT will be higher, which is discussed in Section~\ref{ssec:pw}. To some extent, the active weight of the structure can be increased beyond what is required for survival to obtain the desired gravitational deflection. Thus we need to specify a maximum tolerable gravitational deflection in order to calculate the total weight of the structure.

 \subsection{Compensating for gravitational deflection}
 \label{ssec:cond}
A specification on gravitational deflection has to be arrived at from different considerations, which include the science requirements, the cost of compensating the deflection by re-pointing the small dishes, and the cost of the additional active weight needed to achieve the specified deflection. There are a number of science drivers such as precision pulsar timing, red-shifted HI observations from individual galaxies and HI intensity mapping, which all require excellent system stability and low measurement systematics. These observations need to be done at frequencies below $\sim$2~GHz. Therefore it is desired that the NGAT should operate with no correction for gravitational deflection at frequencies below $\sim$2~GHz. 

As mentioned above, we are assuming that the gravitational deflections can be corrected by re-pointing and focusing the dishes. We consider the case that the diameter of the small dish $d$ is at least $\frac{1}{10}$ of the diameter $D$ of the NGAT structure. Thus at the scale of the dish diameter only a fraction of the full peak-to-peak magnitude of the structural deflection will be effective. This deflection will result in a pointing offset and we would like to keep the change in pointing less than or equal to one tenth of the full width at half power (FWHM) beamwidth of each small dish at frequencies below $\sim$ 2 GHz. At these frequencies we then have,
\bea
    \frac{2 \Delta h_g}{d} & \leq & \frac{1}{10} \frac{1.22 \lambda}{d} \nonumber \\
    \Delta h_g & \leq  & \frac{\lambda}{16} \label{gdef},
\eea
where $\lambda$ is observing wavelength, $\Delta h_g$ is the RMS value of the deflection due to gravity and we consider that the maximum deflection at the scale of $d$ is twice the RMS value. Under these considerations, Eq.~\ref{gdef} essentially states that the wavelength above for which no correction is needed is similar to the gravitational limit of operation for a conventional parabolic reflector telescope. The gravitational limit, $\lambda_g$, is the minimum operating wavelength determined by $\Delta h_g$ for a conventional telescope. If we set this limit to be between 1.5 and 2.5 GHz, then  $\Delta h_g =$ 1.25 -- 0.75 cm.

Let us now consider the operation of NGAT for frequencies above $\sim$ 2 GHz. It will be cost effective to correct for gravitational deflection by moving the feed near the focus rather than moving the whole reflector dish. The shift in pointing in terms of the number of FWHM beamwidths $n$ is given by
\be
    n  =  \frac{\frac{2\Delta h_g}{d}}{1.22 \frac{\lambda}{d}} = 1.6 \frac{\Delta h_g}{\lambda}.
\ee
The maximum desired operating frequency for NGAT is $\sim$30 GHz\cite{b1}. For $\Delta h_g$ in the range 0.75 to 1.25 cm, the shift in beam at 30 GHz corresponds to $n =$ 1.2 -- 2.0. We wish to compensate this shift by moving the feed away from the nominal focus. For a parabolic dish with prime focus operation, the gain changes for the above $n$ values are $<$ 0.5 dB if the focal length over diameter ratio is greater than 0.35 \cite{b4}. Thus the pointing correction to compensate for gravitational deflection can be achieved by just moving the feed at operating frequencies between $\sim$ 2 and 30 GHz.

In summary, we introduce {\em condition 1} for the design of the NGAT -- no pointing correction in the small dishes to compensate for gravitational deflection for operating frequencies below $\sim$ 2 GHz. Below we use 1.5 to 2.5 GHz as the range for this frequency to estimate the telescope weight. 

\subsection{Passive weight of the NGAT}
\label{ssec:pw}
The passive weight for the NGAT will be different from that of a conventional telescope. The passive weight for the NGAT includes the weight of the cooling system for transmitters (weight $\sim$ 600 Kg), helium pump for cryogenic receivers ($\sim$ 400 Kg), electronics and feed support structure ($\sim$ 1000 Kg). Our rough estimate for the weight of these components associated with a small dish is $\sim$ 2 tons. Thus the total excess weight for NGAT is $\sim$ $2N$ tons, where $N$ is the number of small dishes in the NGAT. $N$ is taken approximately as $(D/d)^2$, where $D$ and $d$ are the diameters of the NGAT structure and the small dish respectively.

\subsection{The weight of the NGAT}
\begin{figure}[htbp]
\centerline{\includegraphics[scale=0.58]{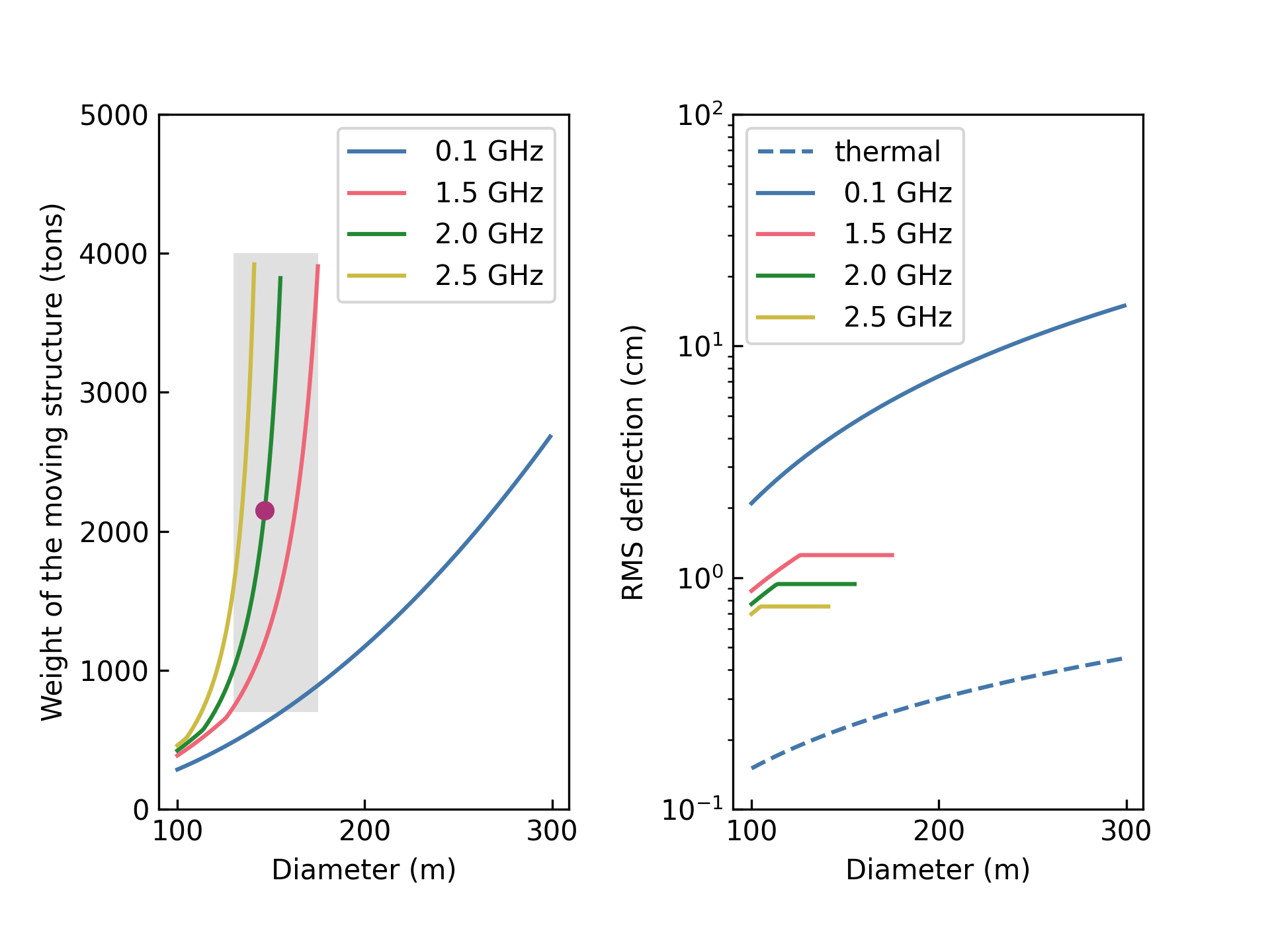}}
\caption{{\bf Left:} Estimated weight of the moving structure in tons as a function of the telescope diameter for the NGAT configuration. The weights are obtained such that the RMS deflection due to gravity is $\leq$ $\lambda_g/16$, where $\lambda_g$ is the wavelength corresponding to the frequencies marked on the legend. For frequencies above those marked, the gravitational deflection needs to be corrected for by re-pointing and focusing the small dishes. The patch shows the range of diameters (130 to 175m) for a NGAT structure which would have a total weight (including the azimuth towers and wheels; see text) less than or equal to that of the GBT. The dot indicates the size and weight of the NGAT-130 (see Section~\ref{sec:ngat130}). The estimated weight in all cases satisfies the survival condition for a maximum wind speed of 136 mph. {\bf Right:} The estimated RMS deflections resulting from structural deformation due to its own weight and temperature gradient of 5\ddeg C for the results shown on the left figure. The dominant deflection is due to gravity. The curves corresponding to frequencies 1.5, 2.0 and 2.5 GHz `flatten' after a certain diameter because the active weight has been increased to maintain the deflection due to gravity at $\lambda_g/16$.}
\label{fig1}
\end{figure}

We consider that the structure is made of steel (see Table 1 of \cite{b2} for a list of the properties of steel). For a given $\lambda_g$ in cm, $D$ in m and $d$ in m, we estimated the following quantities,
\begin{eqnarray}
    w_{ac} & = & 216 (10 \text{cm}/\lambda_g) (D/100 \text{m})^3 + 38 (D/100 \text{m})^2 \\
    w_{ps} & = & 122 (D/100 \text{m})^2 + 2(D/d)^2 \\
    K1 & = & \frac{\lambda_g}{5.3(D/100 \text{m})^2} \\
    K2 & = & 1 + \frac{w_{ps}}{w_{ac}}.
\end{eqnarray}
Here $w_{ac}$ is the active weight in tons, $w_{ps}$ is the passive weight in tons, which includes the additional passive weight due to the components required for the NGAT, K1 and K2 are `passivity factors' (see Eqs. 25, 26, 27 and 8 of \cite{b2}). The diameter $D$ is 1.25 times the diagonal distance of the octahedron since some of the small dishes will be supported by a cantilever. The expressions for $w_{ac}$ are for the survival conditions. If $K1 < K2$, then the active weight needs to be increased to reduce the gravitational deflection so that K2 becomes equal to K1. The total weight of the moving structure of the NGAT is then the sum of the corrected active weight and $w_{ps}$. The results are shown in Fig.~\ref{fig1} for four values of $\lambda_g$ corresponding to frequencies 0.1, 1.5, 2 and 2.5 GHz; the latter three values are consistent with {\em condition 1} discussed in Section~\ref{ssec:cond}. The RMS deflections due to gravity and for an assumed temperature gradient of 5\ddeg C are also shown in Fig.~\ref{fig1}. 

The total weight of the telescope should include the weight of the moving structure plus the towers (supporting the moving structure) and wheels required for the azimuth motion. A detailed design of the towers is required to estimate the weight. In an example telescope design of size 152~m provided by \cite{b2} (see Table V in \cite{b2}), the tower weight is 80\% of the weight of the moving structure. Therefore we take, as a worst case, the total weight of the NGAT as twice the weight shown in Fig.~\ref{fig1}. An upper bound for the weight of a fully steerable NGAT could be taken as the weight of the GBT (8000~tons \cite{b3}). It is indeed possible to build a fully steerable 300~m structure with weight $<$ 8000 tons if we relax {\em condition 1} (see the case for 0.1 GHz in Fig.~\ref{fig1}). But its gravitational deflection and construction cost may be too large to be economical. A lower bound for the size of the NGAT will be determined by the collecting area and transmitter power per dish required for its scientific usage. As discussed below, an NGAT structure with $D > $ 130~m or so will be a very useful scientific instrument. The shaded area in Fig.~\ref{fig1} shows the regions bound by these two limits. In this parameter space, the deflection is dominated by gravitational deformation as seen in Fig.~\ref{fig1}(left). 

\subsection{Operating cost}
\label{sec:power}

\begin{figure}[htbp]
\centerline{\includegraphics[width=1.0\linewidth]{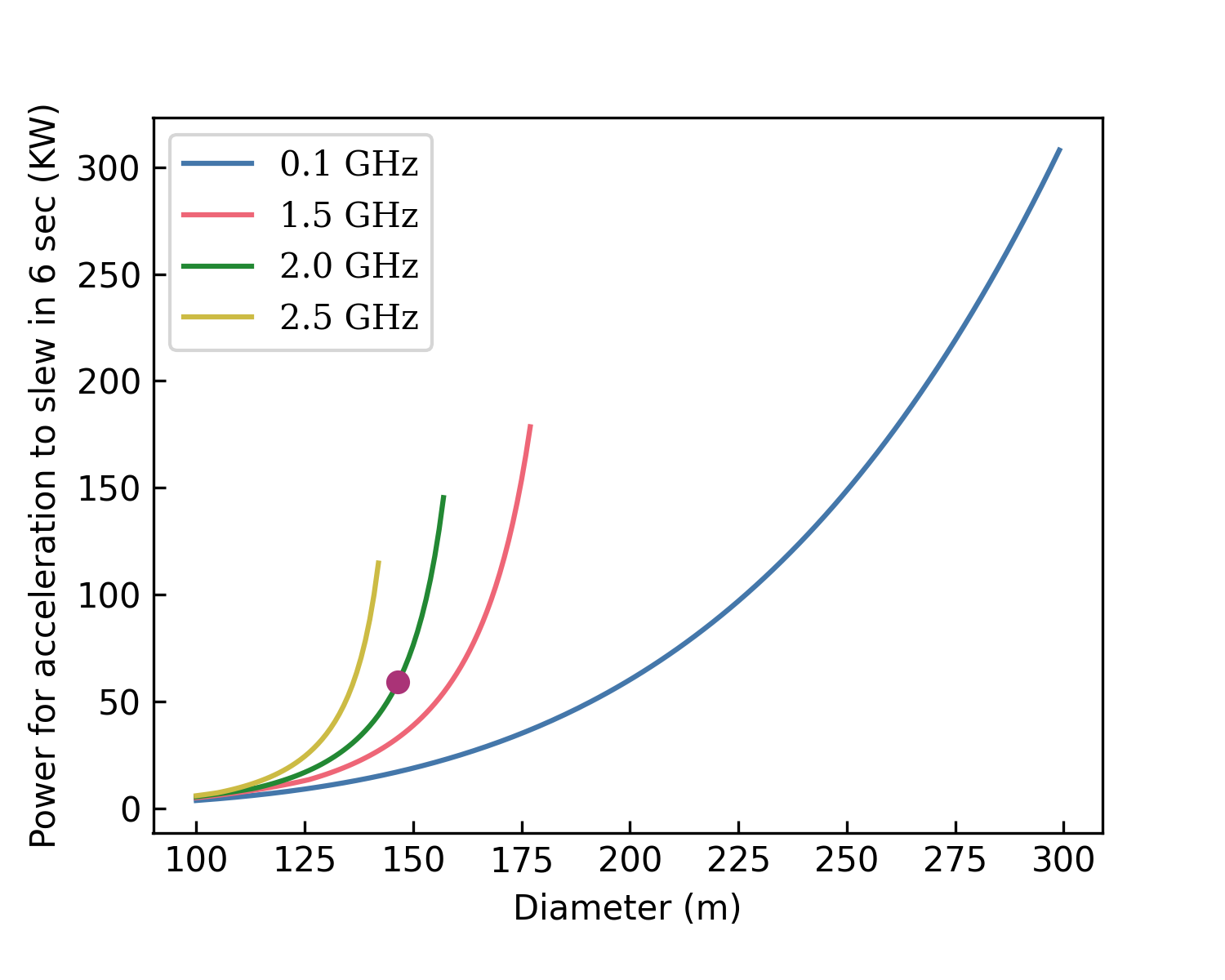}}
\caption{Estimated power consumption to accelerate the NGAT structure from rest to a slew speed of 36 deg/minute at a constant acceleration of 0.1 deg/s$^2$ is plotted as a function of $D$. The dot indicates the power consumption of the NGAT-130 (see Section~\ref{sec:ngat130}), which is about 59 KW.}
\label{fig1.1}
\end{figure}

One of the factors that determine the operating cost is the power required to slew the NGAT structure. Typical slew speed for large telescopes is about 36 deg/minute and acceleration used is about 0.1 deg/s$^2$ (see for example \cite{b3}). We estimate the power consumed to accelerate the NGAT structure from rest to the required slew speed. We approximate the full structure (moving structure + azimuth towers) as a cylinder for this calculation. Fig.~\ref{fig1.1} gives the power consumed to accelerate to the required slew speed in KW as a function of the diameter $D$. As expected, a 300~m NGAT structure will have a higher operating cost (a factor of 5) compared to the NGAT-130 configuration discussed in Section~\ref{sec:ngat130}. 

\section{NGAT-130 and its performance}
\label{sec:ngat130}

\begin{figure}[htbp]
\centerline{\includegraphics[scale=0.3]{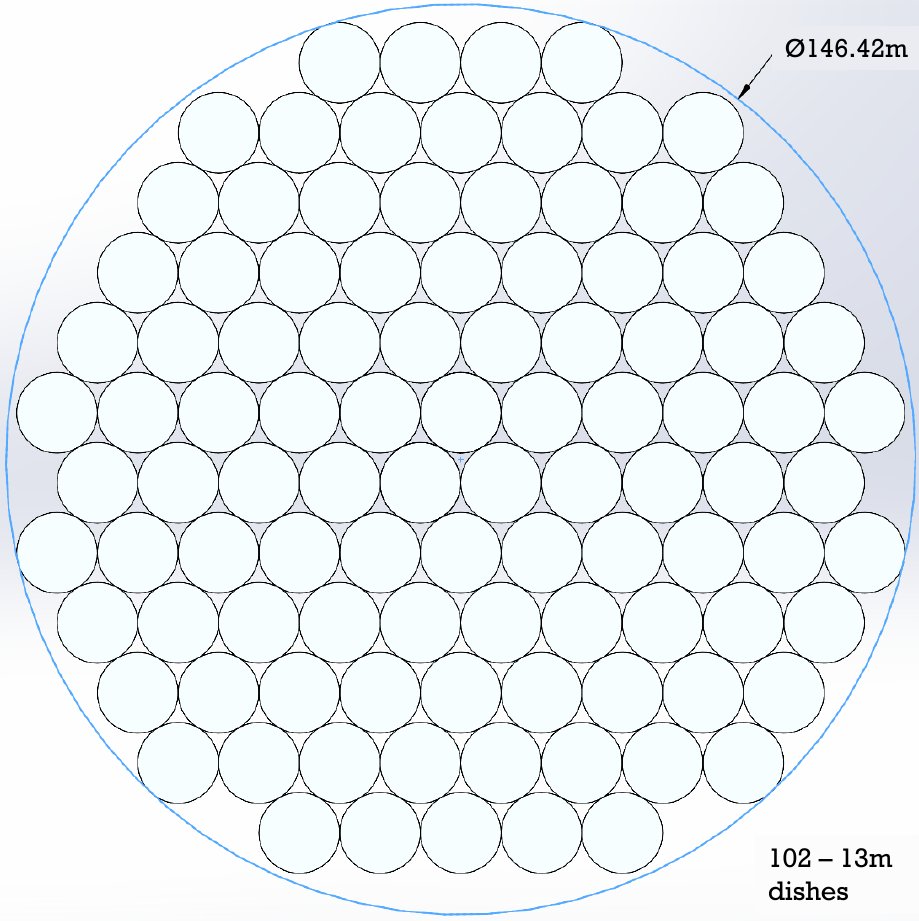}}
\caption{The configuration of the NGAT-130. In this configuration, one hundred and two 13~m diameter dishes form a compact, planar phased array, which will have a collecting area equivalent to a parabolic dish $\sim$130~m in diameter. The size of the telescope structure will be $D \sim$ 146~m.}
\label{fig2}
\end{figure}

\begin{figure}[htbp]
  \centering
  \subfloat[]{%
       \includegraphics[trim={0 0 0 1.0cm}, clip, width=0.9\linewidth]{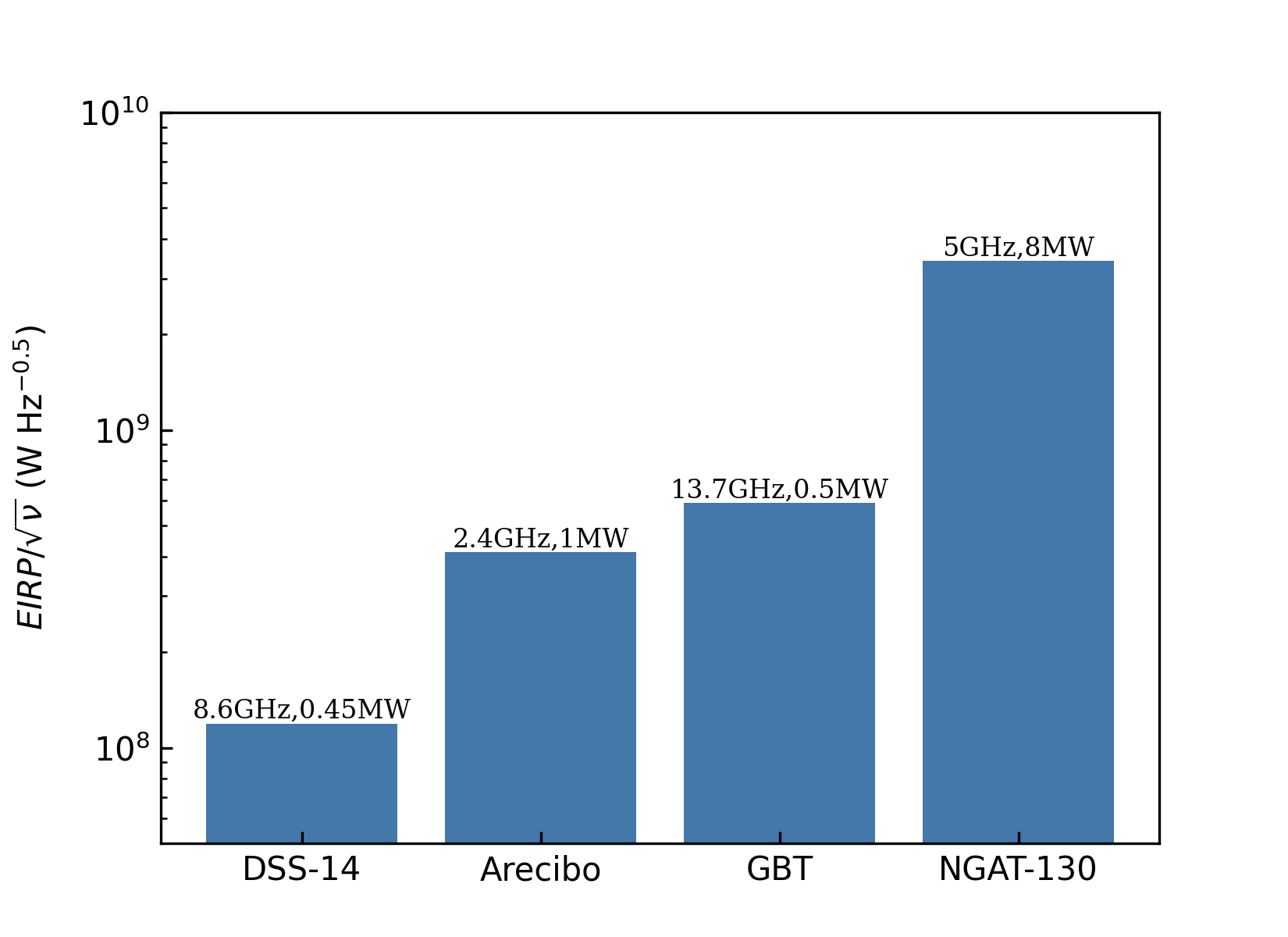}} \\
  \subfloat[]{%
        \includegraphics[trim={0 0 0 1.0cm}, clip, width=0.9\linewidth]{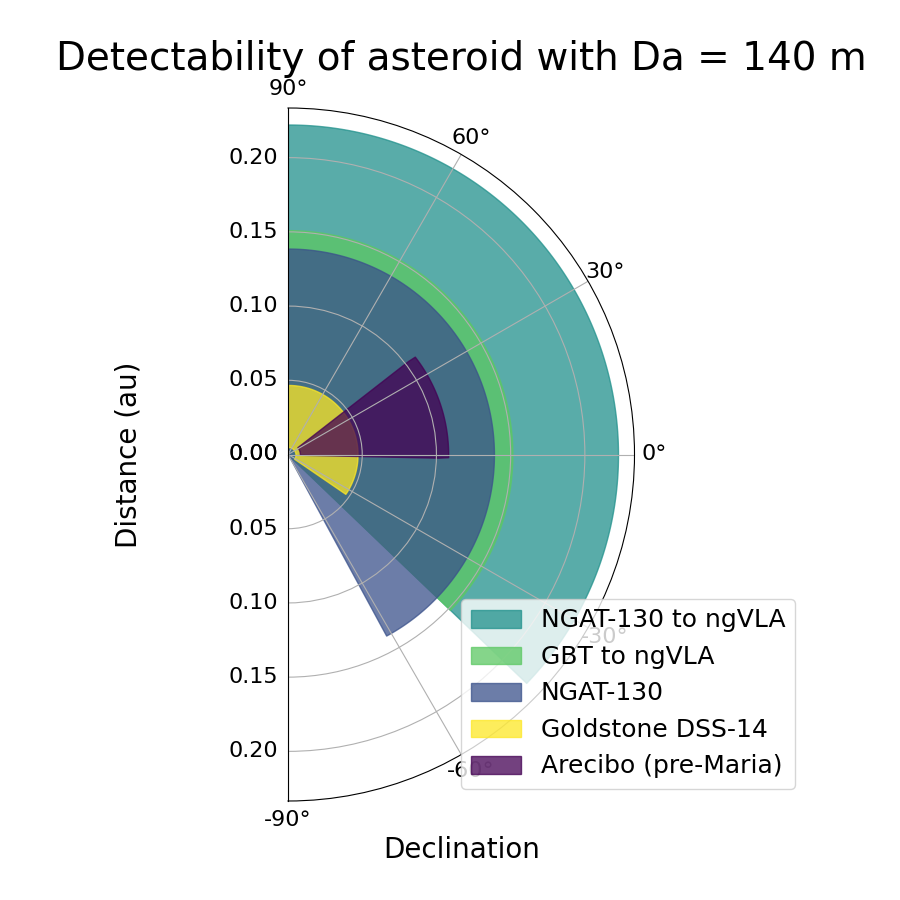}}
\caption{{\bf (a)} The CW transmitter performance of the NGAT-130 compared with other facilities.  {\bf (b)} The distance and declination ranges over which different facilities could detect a potentially hazardous asteroid of size 140~m (with SNR of at least 5 after 30 minutes). }
\label{fig3}
\end{figure}

\begin{figure}[htbp]
  \centering
  \subfloat[]{%
        \includegraphics[trim={0 0 0 1.0cm}, clip, width=0.9\linewidth]{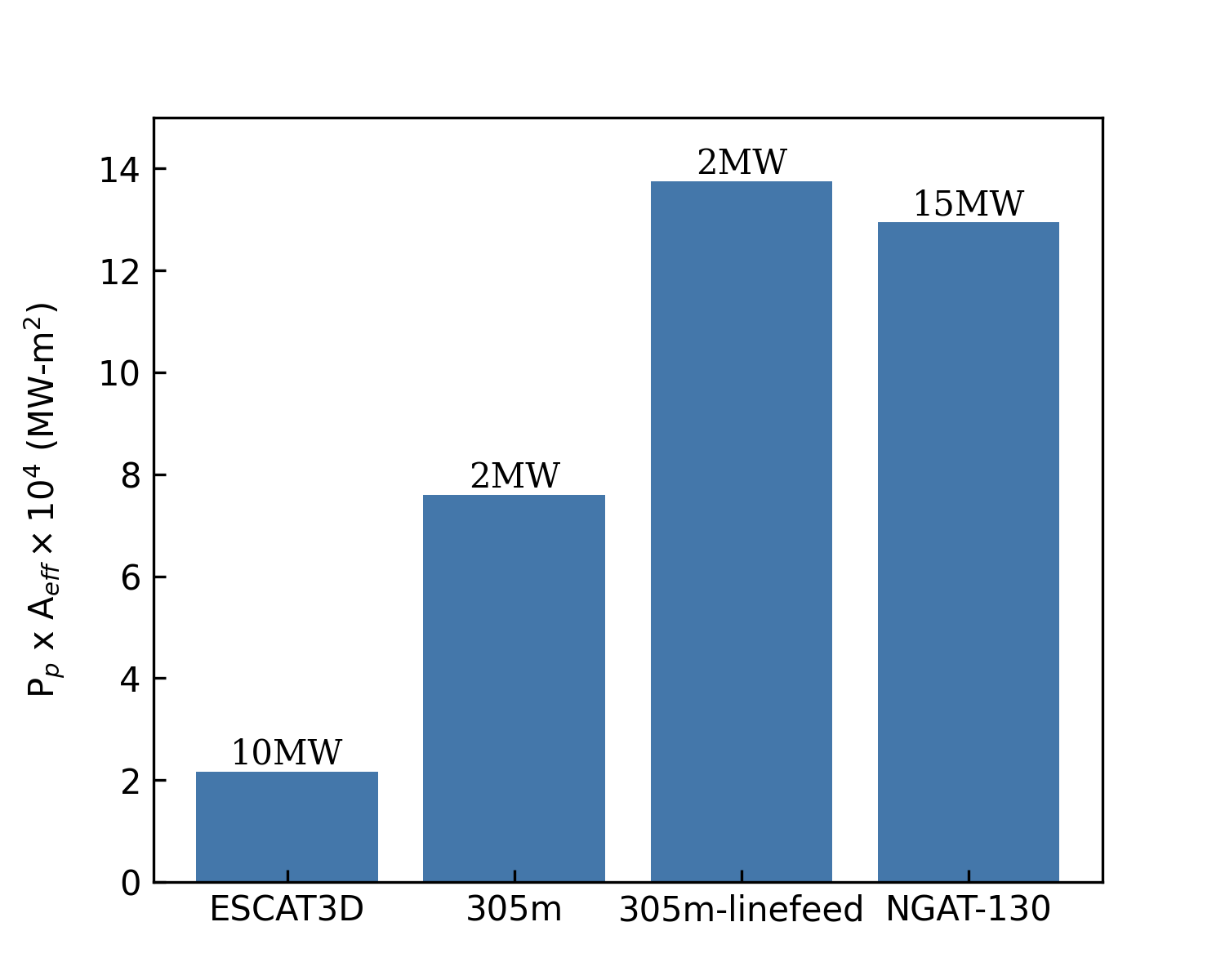}} \\
  \subfloat[]{%
        \includegraphics[trim={0 0 0 1.5cm}, clip,width=0.9\linewidth]{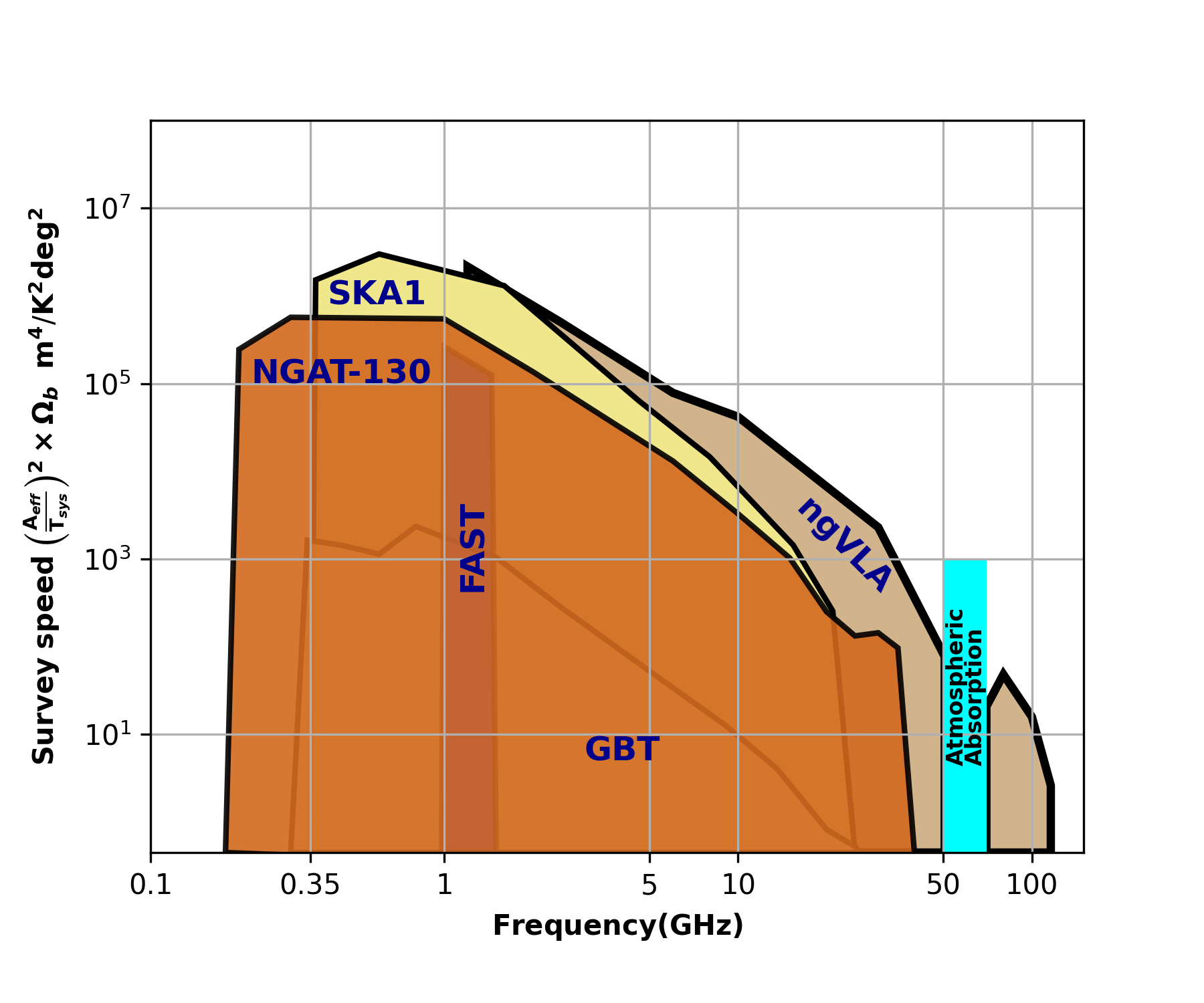}}
\caption{{\bf (a)} The pulsed radar performance of NGAT-130 compared with other facilities. The power aperture product of 305~m telescope illuminated with a line-feed is 6.3\% larger than that of the NGAT-130, but the latter will have larger sky coverage. {\bf (b)} The survey speed of the NGAT-130 compared with other radio telescopes.}
\label{fig4}
\end{figure}

The NGAT would satisfy the requirements of the three existing scientific disciplines at AO. For a given wavelength of operation and a specified system temperature, the ``system factor'' for planetary, atmospheric and radio astronomy sciences are, respectively, $P \times A_{eff}^2$, $P_p \times A_{eff}$ and $A_{eff}$. Here $P$ and $P_p$ are the continuous wave (CW) and peak transmitter powers and $A_{eff}$ is the total effective area of the NGAT. $A_{eff}$ is determined by the number of small dishes (as well as their individual aperture efficiencies), which in turn depends on the size of the telescope structure, and the achievable transmitter power depends on the available electronic components and the details of their implementation (e.g. phased array implementation). The cost model for these two quantities are very different and optimization is definitely required to realize the most economic configuration. Another factor that enters cost optimization is the size of the individual small dish. In the absence of a cost model, we take the power per dish for planetary transmitter as $\sim$80~kW. This value is based on \cite{b5} who discusses the possibility of developing SSPAs (solid state power amplifiers) that could provide such powers. If we consider a dish size of $d = 13$ m, then about 102 dishes could be accommodated within an NGAT structure of size $D \sim 146$ m (see Fig.~\ref{fig2}). This configuration will result in a total CW transmitter power of 8~MW. If the cooling system for this transmitter can be shared with the transmitter for the atmospheric science application, then peak transmitter power of $\sim$150~kW per dish could be used, which provides a net peak power of 15~MW. For astronomical sciences, the collecting area of this NGAT configuration is $\sim$ 70\% larger than that of the GBT. A summary of the telescope configuration, the expected transmitter and receiver performances are given in Table~\ref{tab1}. The total collecting area is equivalent to a parabolic dish of size $\sim$130~m. We compare the performances of the proposed NGAT-130 with other facilities in Fig.~\ref{fig3} \& \ref{fig4}. The total weight (weight of the moving structure + azimuth towers) of the NGAT-130 structure is $\sim$4300~tons for $\lambda_g =$~15~cm, which is 53\% of the weight of the GBT.

\begin{table}[htbp]
\caption{The NGAT-130 configuration}
\begin{center}
\begin{tabular}{|l|r|}
\hline
Small dish diameter & 13~m \\ \hline
Number of dishes & 102 \\ \hline
Diameter of the telescope & 146~m \\ \hline
Total weight of the telescope & 4300~tons \\ \hline
Angular resolution @ 1.4~GHz & 6.2$^{'}$ \\ \hline
Zenith angle coverage & 0\ddeg -- 80\ddeg \\ \hline
Frequency coverage & 0.2 -- 30~GHz \\ \hline
CW transmitter power per dish @ 5~GHz  & 80~kW \\ \hline
Total CW transmitter power @ 5~GHz & 8~MW \\ \hline
Pulsed transmitter peak power per dish @430/220~MHz & 150~kW \\ \hline
Pulsed transmitter net peak power & 15~MW \\ \hline
Receiver temperature \& Aperture efficiency @ 1.4~GHz & 25~K, 0.7 \\ \hline
Receiver temperature \& Aperture efficiency @ 30~GHz & 44~K, 0.6 \\
\hline
\end{tabular}
\label{tab1}
\end{center}
\end{table}

\section{A brief discussion of the science applications}
\label{sec:sci}

A concise list of unique science applications of the NGAT-130 is given below. Overviews of some science applications were given in \cite{b1}, and detailed discussions of the cases below will be presented elsewhere.

\begin{itemize}
\item
    {\em HI intensity mapping:} The dish configuration in the NGAT-130 is similar to the configuration required for experiments like HI intensity mapping \cite{b6}. Unlike currently proposed experiments, the NGAT-130 will be able to track a region in sky which will allow coherent integration of the measured visibility in each of the $uv$ cell. This feature will considerably reduce the integration time required to measure the power spectrum and also will allow repeating such measurements toward different directions in the sky.
    
\item    
    {\em Imaging Gyrosynchrotron emission from Coronal Mass Ejection (CME):} Gyrosynchrotron emission associated with CME is low surface brightness emission and is time variable \cite{b8}. This emission needs to be detected in the presence of strong radio emission from the Sun. Thus imaging Gyrosynchrotron emission requires excellent surface brightness sensitivity, dynamic range, and good instantaneous $uv$ coverage, all of which the NGAT-130 would provide.
    
\item    
    {\em Spectral line surveys:} The large survey speed of the NGAT-130 and spectral baseline stability achievable through the interferometric mode of observations would make it a fantastic instrument to detect, for example, redshifted HI and CO, galactic NH$_3$, and prebiotic molecules over a large sky area.
    
\item     
    {\em Multi-beam phase referenced VLBI:} Precision astrometry using Very Long Baseline Interferometery (VLBI) as well as observations of weak sources at higher frequencies ($>$ 8~GHz) require simultaneous observations of two or more nearby brighter sources to determine the temporal phase variations caused by the atmosphere/ionosphere. The multi-beam capability and higher sensitivity of the NGAT-130 are ideal for such applications.
    
\item     
    {\em Pulsar, Fast Radio Burst (FRB) and NANOGrav Science:} The large field of view and sky coverage combined with the excellent sensitivity make the NGAT-130 a superb facility for pulsar search and characterization, pulsar timing and FRB studies.
    
\item    
    {\em Ionospheric studies and imaging:} The pulsed radar in the NGAT-130 outperforms all other existing facilities. Its large sky coverage will allow the NGAT-130 to make measurements parallel and perpendicular to and also at other angles relative to the earth's magnetic field lines. Further, the NGAT-130 would image the distribution of the scattered radiation from the ionosphere within the primary beam of the 13~m antennas (see for example \cite{b9}). 
    
\item    
    {\em Planetary defense:} The large sky coverage and excellent `system factor' will make the NGAT-130 a unique facility instrument for planetary defense. 
    
\item    
    {\em Space debris:} The large field of view and sensitivity of the NGAT-130 will make it a great instrument for space debris studies (see for example \cite{b10}).
\end{itemize}

\section{Future work}
\label{sec:future}

In this paper we used the von Hoerner model \cite{b2} for large telescope structures, along with a set of assumptions to estimate the size and weight of a realizable and economic structure for the NGAT. Our preliminary study shows that a NGAT of size $\sim$146~m, with one hundred and two 13~m diameter dishes could be developed economically. This configuration, referred to as NGAT-130 (see Table~\ref{tab1}), will have a collecting area equivalent to a single dish of diameter $\sim$130~m. The weight of the structure will be $\sim$4300~tons, which is 53\% of the weight of the GBT. We discussed in Section~\ref{sec:sci} that the NGAT-130 will form a very competitive next generation instrument for the three research fields, planetary, space \& atmospheric and radio astronomy sciences. A number of investigations and development work are required before such a telescope can be realized. These include: (a) develop a mechanical model, estimating its deflections and weight and compare it with the current estimates; (b) investigate if a homologous structure can be developed for the NGAT which can reduce the gravitational deflections; (c) investigate and verify how accurately a finite element model of the structure can predict the structural deformation; (d) investigate the effect of electromagnetic coupling between the dishes and estimate the grating lobe levels (see for example \cite{b11}); (e) develop cost effective 80~kW CW and 150~kW pulsed solid state transmitters; (f) develop a cost effective and high aperture efficiency wideband receiver set for the NGAT; (g) develop pointing and focus correction electronics for compensating the gravitational deflections and prototype them.
We plan to undertake these tasks in the coming years and encourage collaborations to these ends from those who are interested. 




\begin{thebibliography}{00}
\bibitem{b1}  D. Anish Roshi, N. Aponte, E. Araya, H. Arce, L. A. Baker, W. Baan, et al., ``The Future Of The Arecibo Observatory: The Next Generation Arecibo Telescope'', 2021, arXiv:2103.01367, DOI: 10.48550/arXiv.2103.01367.  
\bibitem{b2} S. von Hoerner, ``Design of large steerable antennas'', Astronomical Journal, 1967, vol. 72, pp. 35--47.
\bibitem{b3} R. M. Prestage, K. T. Constantikes, T. R. Hunter, L. J. King, R. J. Lacasse, F. J. Lockman, and R. D. Norrod, ``The Green Bank Telescope'', Proceedings of the IEEE, 2009, Vol. 97, Issue 8, pp. 1382-1390
\bibitem{b4} J. Ruze, ``Lateral-feed displacement in a paraboloid'', ITAP, 1965, Vol. 13, Issue 5, pp. 660-665 
\bibitem{b5} M. S. Net, M. Taylor, V. Vilnrotter, and T. J. W. Lazio, ``A Ground-Based Planetary Radar Array'', 2022, IPN Progress Report, 42-229.
\bibitem{b6} K. Vanderlinde, K. Bandura, L. Belostotski, R. Bond, P. Boyle, J. Brown, et al., ``LRP 2020 Whitepaper: The Canadian Hydrogen Observatory and Radio-transient Detector (CHORD)'', Canadian Long Range Plan for Astronomy and Astrophysics White Papers, LRP2020, 2019, arXiv:1911.01777
\bibitem{b7}  G. Gao, K. Zhang, and S. Sun, ``Optimization of 110 m Aperture Fully Steerable Radio Telescope Prestressed Back Frame Structure Based on a Genetic Algorithm'', Advances in Civil Engineering, 2021, Article ID 3323434, DOI: 10.1155/2021/3323434
\bibitem{b8}T. S. Bastian, and D. E. Gary, ``On the feasibility of imaging coronal mass ejections at radio wavelengths'', Journal of Geophysical Research, 1997, Vol. 102, Issue A7, pp. 14031-14040
\bibitem{b9}J. Stamm, J. Vierinen, J. M. Urco, B. Gustavsson, and Chau, J. L., ``Radar imaging with EISCAT 3D'', Ann. Geophys., 2021, Vol. 39, pp. 119–134. 
\bibitem{b10}J. Murray, and F. Jenet, ``The Arecibo Observatory as an Instrument for Investigating Orbital Debris: Legacy and Next Generation Performance'',  Planetary science journal, 2022, Vol. 3, p. 52, DOI: 10.3847/PSJ/ac4d96 
\bibitem{b11} J. Hyatt, J. Berkson, Z. Hatfield, D. W. Kim, N. Nguyen, D. A. Roshi, S. Dharmalingam, M. Sulzer, ``Grating lobe suppression for the next generation Arecibo Telescope concept'', Proceedings of the SPIE, 2021, Vol. 12078, DOI: 10.1117/12.2603637
\end{thebibliography}
\end{document}